  \documentstyle[12pt]{article}
  
\catcode`@=11
\def\secteqno{\@addtoreset{equation}{section}%
\def\theequation{\thesection.\arabic{equation}}}
\catcode`@=12
\secteqno
\newcommand{\be}{\begin{equation}}
\newcommand{\ee}{\end{equation}}
\newcommand{\bea}{\begin{eqnarray}}
\newcommand{\eea}{\end{eqnarray}}
\newcommand{\bref}[1]{(\ref{#1})}
\newcommand{\nn}{\nonumber}

\newcommand{\A}{\alpha} \newcommand{\B}{\beta} \newcommand{\gam}{\gamma} 
\newcommand{\G}{\Gamma} \newcommand{\D}{\delta}
 
\newcommand{\T}{\theta} 
   
         \newcommand{\lam}{\lambda}
           \newcommand{\s}{\sigma}
          \newcommand{\w}{\omega}
\newcommand{\z}{\zeta}          \newcommand{\x}{\xi}
\newcommand{\h}{\eta}           
           
\newcommand{\W}{\Omega}

\newcommand{\Tb}{{\overline\theta}}
\def\pa{\partial}

\def\CG{{\cal G}}\def\CL{{\cal L}}
\def\CF{{\cal F}}

\def\t{\tilde}
\def\l{{\ell}}

\def\vsej{\vskip 4mm} 
\newcommand{\slPi}{/ {\hskip-0.27cm{\Pi}}}
\begin{document}
\vfill
\vbox{
\hfill  March 26, 1999 \null\par
\hfill TOHO-FP-9962\null\par 
\hfill TIT/HEP-418\null\par
\hfill }\null
\vskip 20mm
\begin{center}
{\Large\bf SO(2,1) Covariant IIB Superalgebra}\par
\vskip 10mm
{\large Mitsuko\ Abe,\ Machiko\ Hatsuda$^\dagger$,}\par
{\large Kiyoshi\ Kamimura$^{\dagger\dagger}$
\ and\ Takashi\ Tokunaga$^{\dagger\dagger}$}\par
\medskip
{\it Department of Physics, Tokyo Institute of Technology, 
Meguro, Tokyo, 152-0033, Japan\\
$^{\dagger}$ Department of Radiological Sciences,
Ibaraki Prefectural University of Health Sciences,\\
\ Ami\ Inashiki-gun\ Ibaraki\ 300-0394, Japan \\
$^{\dagger\dagger}$ Department of Physics, Toho University, 
Funabashi\ 274-8510, Japan
}\par
\medskip
\end{center}
\vskip 20mm
\begin{abstract}
We propose a type IIB super-Poincar${\acute{\rm e}}$ 
algebra with SO(2,1) covariant central extension. 
Together with SO(2,1) and SO(9,1) generators,
 a SO(2,1) triplet (momenta), a Majorana-spinor doublet (supercharges)
and a Rarita-Schwinger central charge generate a group, $G$.
We consider a coset $G/H$ where $H$=(SO(2)$\times$Lorentz), 
and the SL(2,R) 2-form doublet is obtained by the coset 
construction. 
It is shown that U(1) connections, whose strengths are associated with 
2-forms, are recognized as coordinates of the enlarged space.
We suggest that this is the fundamental algebra governing the superstring 
theories which explains the IIB SL(2,R) duality and geometrical origin of U(1) 
fields.
\end{abstract} 
\par
\newpage
\setcounter{page}{1}
\parskip=7pt
\section{ Introduction}\par
\indent

Origin of the SL(2,Z) duality of type IIB supergravity theory 
\cite{SchSG,Shsl} is still an unsolved problem. 
Possible geometrical origins of the IIB SL(2,Z) duality in D-brane 
theories are examined in higher dimensional theories, 
such as 12-dimensional theory (F theory \cite{Vafa}) and 
13 or 14-dimensional theories \cite{Bars}.
On the other hand, it is known that a hidden dimension (11-th direction) 
appears as worldvolume U(1) fields on a D2-brane \cite{Twn,SchDD}.
It is also expected that two hidden dimensions (11- and 12-th directions) 
are described as two transverse components of worldvolume U(1) gauge 
fields on a D3-brane \cite{Tsy}.
 Towards the higher dimensional theories the worldvolume U(1) fields will 
play important roles in exploration of the geometrical origin of the IIB SL(2,Z) duality.

In order to find better understanding on this point, 
the superalgebra and the Hamiltonian analyses are proper tools. 
The superalgebras (SUSY algebras) for Dp-branes contain central charges 
representing their NS/NS and R/R gauge couplings \cite{pol,Hull}.
SUSY central charges, in general, arise from the Wess-Zumino terms or 
topological terms \cite{azc}.
The Wess-Zumino terms or topological terms in the Dp-brane theories
come from NS/NS and R/R gauge couplings \cite{Doug}
where the NS/NS and R/R gauge fields appear as representations of the 
type IIB SL(2,R) symmetry (SL(2,Z) at the quantum level). 
Concrete expressions of SUSY central charges for Dp-branes 
are given as the p-form brane charges \cite{azc} and 
others depending on the worldvolume U(1) field \cite{HK,Ham,KH,MK}.
For a D1-brane the worldvolume electric field corresponds to the NS/NS charge 
and the D1-brane charge corresponds to the R/R charge \cite{HK}.
For a D3 brane, the electric field and the magnetic field appear 
in the NS/NS and the R/R central charges respectively in its superalgebra,
and in the Hamiltonian its BPS mass is obtained as the electro-magnetic energy
\cite{KH,IIK3}.
The analyses of the superalgebra in the Hamiltonian formalism explain 
the BPS mass property of Dp-branes as a consequence of worldvolume U(1)
 excitations \cite{HK,KH,IIK3}. 
It is appropriate to examine the IIB D-brane superalgebra 
to explore the geometrical origin of the worldvolume U(1) gauge fields 
and the IIB SL(2,Z) duality.   
\vsej

Siegel have constructed the Wess-Zumino term of N=1 fundamental 
string \cite{Siegel} using a central extension of super-Poincar$\acute{\rm e}$ 
algebra proposed by Green \cite{Green}. 
The Wess-Zumino term is given in a bi-linear combination of invariant forms
and the Lagrangian density is invariant under super transformations by itself. 
Recently Sakaguchi \cite{Sakag} discussed central extension of 
superalgebras inspired by works in \cite{Sez} for p-branes. 
These extended superalgebras contain new bosonic generators corresponding to
brane charges and new fermionic generators
in addition to original super-Poincar$\acute{\rm e}$ generators.
Using with the algebras and the coset construction, he discussed the IIA, IIB
and (p,q) string actions.

It is important to notice that the type IIB D-brane superalgebra naturally 
fits with the SL(2,R) covariantization \cite{Bars}. The total momentum  
together with two vector central charges associated with two second rank 
anti-symmetric fields $B_{\mu\nu}$ form a  SL(2,R) triplet.
In this paper we propose an extension of the super-Poincar$\acute{\rm e}$ 
algebra manifestly covariant under SO(2,1) $\cong$ SL(2,R) :
\bea
\{Q_{A\A},~Q_{B\B}\}&=&2i~P_{jm}~(C\G^m)_{\A\B}~(c\varrho^j)_{AB},
\label{QQP}
\\
\left[P_{im},~Q_{A\A}\right]&=&i~Z_{iA\B}{(\G_m)^\B}_\A,
\label{PQZ}
\\
\left[P_{im},P_{jn}\right]&=&0.
\eea
Here $Q_A$ is the SO(2,1) doublet ``supercharges'', $P_j$ is the SO(2,1) 
triplet  ``momenta" operators and
$Z_{iA}$ is the SO(2,1) spin 3/2 fermionic  central charges. 
It will be discussed  the identification of the SO(2,1) with
the SL(2,R) symmetry of the type IIB supergravity theory.
We suggest that this is the basic symmetry algebra of the superstring and 
the super D-brane theories.{\footnote{It could appear 
higher rank tensor charge in \bref{QQP} \cite{Bars}.}}

The organization of this paper is the following:
In section 2, we propose the SO(2,1) covariant IIB SUSY algebra and 
examine a condition from the Jacobi identity. 
In section 3, the coset construction is applied to find 
invariant forms.
In section 4, the type IIB SL(2,R) representations are given in terms of 
SO(2,1)/SO(2) coset representations. It is also shown the identification
of the SO(2,1) with the type IIB SL(2,R).
In section 5, 
the SO(2,1) invariant D-string action is considered in terms of the
invariant forms. Especially we show how the U(1) fields are
described in terms of coordinates in the enlarged space.
Some discussions are given in the last section. \par
\vsej
\section{SO(2,1) covariant central extension of Superalgebra}
\indent

SO(2,1), which is isomorphic to SL(2,R)  and SU(1,1),
is the Lorentz group in 2+1 dimensions.  
The generators $N_{ij},~(i,j=\hat{0},\hat{1},\hat{2})~$ are satisfying
the Lorentz algebra
\bea
\left[N_{ij},N_{k\l}\right]&=&i~
(\h_{\l i}~N_{jk}-\h_{\l j}~N_{ik}
-\h_{ki}~N_{j\l}+\h_{kj}~N_{i\l}),
\eea
where $\eta_{ij}={\rm diag}(-1,1,1).$
The IIB supersymmetry generators $~Q_{A}~(A=1,2)~$ are 10D Majorana-Weyl 
spinors with the same chirality. We assign them as a 
SO(2,1) Majorana spin 1/2 doublet,{\footnote
{ 
The gamma matrices in 2+1 dimensions are satisfying $
\left\{\varrho^{i},\varrho^{j}\right\}=2\eta^{ij}$. 
In the Majorana representation 
\bea
\varrho^{\hat{0}}=-i\tau_2~,
\varrho^{\hat{1}}=\tau_1~,~
\varrho^{\hat{2}}=-\tau_3~~~~~{\rm and}~~~~~
c\varrho^{\hat{0}}={\bf 1}~,
c\varrho^{\hat{1}}=\tau_3~,~
c\varrho^{\hat{2}}=\tau_1,~~
\eea
where the charge conjugation matrix is 
$~c=i\tau_2~,~(\varrho^{i})^t=-c~\varrho^{i}~c^{-1}$.
}}
\bea
\left[Q_A,~N_{jk}\right]&=&-\frac{i}{2}~Q_B(\varrho_{jk})^B_{\ A},~~~~~~~
\varrho_{jk}\equiv\frac12\varrho_{[j}\varrho_{k]}.
\eea
where $(i/2)\varrho_{jk}$ is the 
spinor representation of $N_{jk}$.
As a consequence three 10D vector charges $P_{i}$ appearing in the 
superalgebra \bref{QQP},
form a  SO(2,1) vector triplet,
\bea
\left[P_{i},~N_{jk}\right]&=&i~
(\eta_{ij}P_{k}-\eta_{ik}P_{j}).
\eea
The fermionic central charges  {\footnote  
{We call  $~Z_{iA\B}~$'s ``central charges" in the sense that
they (anti-)commute with $P,Q$ and $Z$, {\it i.e.}~
$\{Z,Z\}=\{Z,Q\}=[Z,P]=0$.}} 
$~Z_{iA\A}~$ appearing in the $P,Q$ commutator \bref{PQZ} 
transform as spinor-vector under SO(2,1),
\bea
\left[Z_{iA\A},~N_{jk}\right]&=&i~
\eta_{i[j}Z_{k]A\A}~-\frac{i}{2}~Z_{iB\A}(\varrho_{jk})^B_{\ A}~.
\eea
10D Lorentz transformation generators are $M_{mn}$,
and $P_{i}$'s transform as vectors 
and $Q_{A}$'s and $Z_{iA}$'s as Majorana-Weyl spinors.

Jacobi identities are satisfied trivially except one of three $Q$'s,

\noindent
$\sum_{cyclic}\left[\{Q_{A\A},Q_{B\B}\},Q_{C,\gam}\right]~=0$.
It requires, using the 10 dimensional cyclic identity,
\bea
Z_{jB\A}(\varrho^j)_{\ A}^{B}~&=&0
\label{ZREL}
\eea
which tells $Z_{jB}$ is SO(2,1) irreducible spin 3/2 generator.
With this irreducibility condition  $(Q_{A\A},P_{im},M_{mn},
Z_{iA\A},N_{ij})$ form a 
closed algebra $\CG$. \par

The algebra $\CG$ is a SO(2,1) covariant generalization of ones 
recently proposed by Sakaguchi \cite{Sakag}. Actually if one of
fermionic generators $Z_i$ is eliminated explicitly by using \bref{ZREL} 
it reproduces algebras in ref.\cite{Sakag}. 

\vsej
\section{ Coset construction}
\indent

In this section we construct invariant forms 
using the nonlinear realization of
the group $G$ of the algebra ${\cal G}$ introduced in the last section.

The coset we are going to use is ${G}/{H}$.
The numerator is the SO(2,1) covariant IIB super-Poincar$\acute{\rm e}$ 
group with the central extension introduced in previous sections. 
The subgroup $H$ is product of the homogeneous Lorentz group and SO(2) which
is a subgroup of SO(2,1). 
We parameterize the coset in the following form
\bea
g&=&g_N~g_Z(\xi)~g_P(y)~g_Q(\T),
\label{gpare}\\
\nn\\
&&
\left\{\begin{array}{l}
g_Z~=~e^{-iZ_{iA\A}\x^{iA\A}}=~e^{-iZ\x}~
\\
g_P~=~e^{iP_{im}~y^{im}}~~~=~e^{iPy}
\\
g_Q~=~e^{-iQ_{A\A}\T^{A\A}}~~=~e^{-iQ\T}
\end{array}\right.
\label{gda}
\eea
The SO(2,1)/SO(2) part $g_N$ is usually parameterized by two scalars, 
dilaton and axion.
Due to the irreducibility condition of $Z_{iA}$ \bref{ZREL} the coordinates
$\xi^{iA}$ are not independent and $g$ is invariant under
\bea
\delta \xi^{jA\A}&=&(\varrho^j)^A_{~B} \Lambda^{B\A}. 
\label{xigauge}
\eea
The transformation of the coset element $g$ under the group $G$ is
\bea
g&\to&\Lambda~g~h^{-1},~~~~~~~~~~~\Lambda\in G,~~h\in H,
\eea
where $h$ is (induced local) subgroup transformation parameterized as
\bea
h&=&e^{\frac{i}{2}M_{mn}\w^{mn}}~e^{i N_{\hat{1}\hat{2}} \psi}.
\eea
It determines the transformation of the coset coordinates.
Under SO(2,1) transformations 
$ y^{\hat{0}m}$ is a scalar,
$ y^{{\underline{i}} m},({\underline{i}}=\hat{1},\hat{2})$ 
is SO(2) vector doublet{
\footnote{In the adjoint representation $h$ 
represents a rotation of three vectors around the $\hat{0}$ th axis.}},
$ \T^{A\A}$ and $ \xi^{\hat{0}A\A}~(A=1,2)$ are SO(2) spinor doublets,
$ \xi^{{\underline{i}} A\A}$ is SO(2) vector-spinor. 
It is noted that they are not transformed as SO(2,1) multiplets but of SO(2)
due to the parameterization \bref{gpare}. \par

\medskip

Left invariant one form is constructed as
\bea
\W&\equiv&-ig^{-1}dg~=~\frac12 N_{ij}L_N^{ij}~+~
QL_Q~+~PL_P~+~ZL_Z.
\eea
Under the transformation of $G$ the M-C form transforms as
\bea
\Omega&~\to&~~h~\Omega~h^{-1}-ihdh^{-1},~~~~~~~~h\in H.~~
\eea
$L^{\hat{1}\hat{2}}_N$ transforms as the SO(2) gauge connection while other
one forms $L$'s transform homogeneously as 
SO(2) covariant quantities. 
That is $ L_{P}^{\hat{0}m}$ is a SO(2) scalar,
$ L_{P}^{{\underline{i}} m},({\underline{i}}=\hat{1},\hat{2})$ is SO(2) 
vector doublet,
$ L_Q^{A\A}$ and $L_Z^{\hat{0}A\A},~(A=1,2)$ are SO(2) spinor doublets
and $ L_Z^{{\underline{i}} A\A}$ is SO(2) vector-spinor.
The one form coefficients are
\bea
L_Q^{A\A}&=&D{\T}^{A\A},~
\nn\\
L_{P}^{im}&=&D y^{im}~+~\bar{\theta}\Gamma^m(c\varrho^i)D\theta, 
\nn\\
L_Z^{iA\A}&=&D \x^{iA\A}~+~(\G^m\T)^{A\A}
\{~D y^i_m~+~\frac{1}{3} ({\overline{\T}}\G_m~(c\varrho^i)~D\T)\}.
\label{MCone}
\eea
where $D$ is the SO(2) covariant derivative
in which $-ig_Ndg_N^{-1}$ plays a role of the gauge connection.
The M-C equations,~$d\W~+~i\W^2~=~0~,$
holds as
\bea
DL_Q&=&0,
\\
DL^i_P- ({\overline L_Q}\G(c\varrho^i)L_Q)&=&0,
\\
Z_i(~DL^i_Z-\G L_Q~L^i_P~)&=&0.
\label{MCeqb}
\eea
In the last one $Z$ is kept multiplying since $Z$'s are not independent,
\bref{ZREL}, and  there is an ambiguity in defining $L_Z$'s.
\medskip

Next we use the M-C one forms 
\bref{MCone} to obtain bosonic two forms which satisfy two requirements:
 One is that the new coordinates $y^{\underline{i}}$ and $\xi^{iA}$
appear only in exact forms \cite{Siegel}.
The other is the invariance under \bref{xigauge}.
From now on we assume constant dilaton and axion
 and the SO(2) covariant derivatives are replaced by the ordinary derivatives.
There exists only three such two forms. 
One of them is an exact form 
\bea
\CF^{12}&\equiv&\frac12[~L_P^{\hat{1}}~L_P^{{\hat{2}} }~-~
{\overline L}_Q
\{c\varrho^{\hat{1}}(L_Z^{{\hat{2}}}-\varrho^{{\hat{0}}}\varrho^{{\hat{2}}}
L_Z^{{\hat{0}}})~-~
c\varrho^{\hat{2}}(L_Z^{{\hat{1}}}-\varrho^{{\hat{0}}}\varrho^{{\hat{1}}}
L_Z^{{\hat{0}}})\}].
\eea
Other two are $\CF^{{\underline{i}} },
({{\underline{i}}}=\hat 1,\hat 2)$ defined as
\bea
\CF^{{\underline{i}} }&\equiv&\frac12[~L_P^{\hat{0}}~L_P^{{\underline{i}} }~-~
{\overline L}_Q
(c\varrho^{\hat{0}}L_Z^{{\underline{i}} }+c\varrho^{{\underline{i}} }~
L_Z^{\hat{0}})]
\nn\\
&=&\frac12[~dx~dy^{{\underline{i}} }~-~d\Tb~d(
c\varrho^{\hat{0}}\x^{{\underline{i}} }+
c\varrho^{{\underline{i}} }\x^{\hat{0}})]~-~
(\Tb\G ~c\varrho^{{\underline{i}} }~d\T)(dx+\frac12\Tb\G d\T)~
\nn\\
& = & dA^{\underline{i}}~-~B^{\underline{i}},
\label{ninv2f}
\eea
where
\bea
B^{\underline{i}} &\equiv&(\Tb\G c\varrho^{\underline{i}} d\T)(dx+\frac12\Tb\G d\T),
\eea
and
\bea
 F^{\underline{i}}&\equiv&dA^{\underline{i}}~=~
d[-\frac12\{~y^{\underline{i}}~dx~-~(\overline\x^{\underline{i}}
c\varrho^{\hat{0}} 
+\overline\x^{\hat{0}}c\varrho^{\underline{i}} )d\T \}].
\label{BBFF}
\eea
$B^{\underline{i}}~({\underline{i}}=\hat 1,\hat 2)$ are pullbacks of the NS-NS 
and R-R two forms of the flat background \cite{SchDD}.
$A^{\underline{i}} ({\underline{i}}=1,2)$ are corresponding to the DBI
world volume U(1) potential and its counterpart in the R-R sector.
In this way the world volume U(1) potentials acquire
an interpretation in terms of coordinates in the extended 
superspace \cite{Sakag}.
$\CF^{{\underline{i}} }$, $A^{{\underline{i}} }$ and  
$B^{{\underline{i}} }$ transform as SO(2) vector 
doublets under G. 
\par
\vsej

\section{ IIB SL(2,R) multiplet}
\indent

Next we use the M-C one forms 
\bref{MCone} to describe the type IIB multiplet 
and examine their relation to the NS-NS and R-R two forms
following SL(2,R) $\cong$ SO(2,1) transformation rules.\par
\medskip

The SL(2,R) representations of the type IIB supergravity multiplet are
the followings: the metric and the antisymmetric rank four tensor fields are
singlet and 
the dilaton $\phi$ and axion $\chi$
 are coordinates in the nonlinear realization transforming as
\bea
\mu
=e^{\phi}\pmatrix{\chi^2+e^{-2\phi}&\chi\cr \chi&1}~~~,~~\mu\rightarrow
\Lambda \mu \Lambda^t~~,~~\Lambda \in {\rm SL(2,R)}~~.\label{dilaton}
\eea
Two kinds of second rank antisymmetric tensor fields ${\cal B}=
(B^{NS},B^{R})$ are SL(2,R)
doublet  
\bea
{\cal B}~\rightarrow (\Lambda^t)^{-1} {\cal B}~~.\label{2form}
\eea

Since $y^{\hat{0}m}$ is a 10D vector coordinate invariant under SO(2,1), we  
identify $y^{\hat{0}m}$ to the position coordinate of branes $x^m$
and the SO(2,1) and SUSY invariant one forms are
\bea
L_P^{\hat{0}m}&=&dy^{\hat{0}m}+\Tb\G^m c\varrho^{\hat{0}}d\T~
=~dx^m~+\Tb\G^md\T~\equiv~\Pi^m.
\label{LP}\eea
The induced metric of the branes is 
\bea
G_{\mu\nu}&=&\h_{mn}(\Pi_{\mu})^m(\Pi_{\nu})^n,~~~~~~~~~
\Pi^m~=~d\s^\mu(\Pi_{\mu})^m
\eea
and is the SO(2,1) invariant Einstein metric. The $\Pi^m$ and 
\bea
L^{A\A}_Q&=&d\T^{A\A}
\label{LQ}
\eea
are SUSY invariant one forms in the usual space.
The latter transforms as a SO(2) spinor doublet under the SO(2,1).

In order to find the relation between the SO(2,1) considered above
and the type IIB SL(2,R), we examine the transformation rules by 
using concrete expression of the coset SO(2,1)/SO(2).
It is convenient to use the SL(2,R) algebra $L_n$ 
\bea
\left[L_n,L_m\right]&=&(n-m)L_{n+m}~,~(n,m=-1,0,1)~.
\eea
rather than the SO(2,1) 
algebra $N_{ij}$. They are relating  by
\bea
L_{+1}&=& -i(N_{\hat{0}\hat{2}}+ N_{\hat{1}\hat{2}})~,~
L_{-1}~=~ i(N_{\hat{0}\hat{2}}- N_{\hat{1}\hat{2}})~,~
L_0=iN_{\hat{0}\hat{1}}~~.
\eea
We parameterize the coset element $g_N\in$ SO(2,1)/SO(2) 
also by introducing redundant SO(2) gauge degrees of freedom $\varphi$ 
\cite{SchSG} as
\bea
g_N=e^{ L_{+1}\chi}e^{ L_0\phi}e^{iN_{\hat{1}\hat{2}}\varphi}
\eea
and it transforms under SL(2,R) $\cong$ SO(2,1) as
\bea
g_N~~\to~~\Lambda~
g_N~h^{-1},~~~~~~~~\Lambda\in SL(2,R),~~h\in SO(2).
\label{Ktrans}
\eea
The SO(2) transformation $h$ is determined up to that of $\varphi$.
The $2\times 2$ matrix representation of $g_N$ is written as 
{\footnote{ 
In the matrix representation  
\bea
L_{+1}=\pmatrix{0&1\cr 0&0}~~,~~
L_{-1}=\pmatrix{0&0\cr -1&0}~~,~~
L_{0}=-\frac{1}{2}\pmatrix{1&0\cr 0&-1}.
\eea
}}
\bea
g_N \vert_{2\times 2}=K=VR_{\varphi}~,~
\left\{\begin{array}{l}
V=\pmatrix{1&\chi\cr0 &1}~\pmatrix{e^{-\frac{\phi}{2}}&0\cr 0&
e^{\frac{\phi}{2}}}=
\pmatrix{e^{-\frac{\phi}{2}}&e^{\frac{\phi}{2}}\chi\cr0 &
e^{\frac{\phi}{2}} }\\
R_{\varphi}=\pmatrix{\cos\frac{\varphi}{2}&
-\sin \frac{\varphi}{2}\cr\sin\frac{\varphi}{2} 
&\cos\frac{\varphi}{2}}
\end{array}\right.~
.\label{gN}
\eea
These coset coordinates $\phi$ and $\chi$ are identified with
the ones of the IIB supergravity; dilaton and axion, since
the bi-linear SO(2) invariant combination of $K$ \bref{gN}
gives $\mu$ in \bref{dilaton}
\bea
K K^{t}&=&VV^{t}~=~\mu
\eea  
and it transforms in the same manner as \bref{dilaton}.

This $2\times 2$ matrix representation $K$ of the SO(2,1)/SO(2) 
makes SO(2) {\it spinor} doublets to be SO(2,1)  doublets, e.g. 
$\t\T^A=K^A_{~B}\T^B$ transforms as $\t\T\to\Lambda\t\T$.
In order to define SL(2,R) doublet two forms ${\cal B}^A$ from SO(2)
{\it vector} doublet ${B^{\underline{j}}}$ in \bref{BBFF} 
it is necessary to convert the vector index of ${B^{\underline{j}}}$ to 
the spinor index. 
It is possible if there exists an SO(2) spinor
$\Psi=(\sin{\t\varphi/2},\cos{\t\varphi/2})$. We define
\bea
{K^{(v)}}^A_{~~{\underline{j}}}&\equiv&
K^A_{~B}(\varrho_{\underline{j}})^{B}_{~C}\Psi^C~=~(V(\phi,\chi)
R_{\varphi}R_{\t\varphi})^A_{~~{\underline{j}}}.
\label{Kvec}
\eea
Then the SO(2) vector doublet of two forms 
\bref{BBFF} can be lifted to SO(2,1) doublet 
\bea
{\cal B}^A&=&(K^{(v)})^{A}_{~~ \underline{j} }
B^{\underline{j}}.
\eea
We can take the SO(2) gauge degrees of freedom $\varphi$ satisfying
$\varphi+\tilde{\varphi}=0$, 
so that the SO(2,1)/SO(2) coset element $K^{(v)}$ takes a conventional 
form $V(\phi,\chi)$ expressed in terms of a dilaton and an axion,
\bea
{\cal B}^A&=&V(\phi,\chi)^{A}_{~~ \underline{j} }B^{\underline{j}},~~~~~~~~
V=\pmatrix{e^{-\frac{\phi}{2}}&e^{\frac{\phi}{2}}\chi\cr0 &
e^{\frac{\phi}{2}} }.
\eea

\vsej

\section{IIB Dp-branes}
\indent

In this section we consider SL(2,R) invariant Dp-branes. 
The action have been discussed in  SL(2,R) covariant forms in
\cite{Towc}. 
$H$ invariant Lagrangians which are constructed from 
$L_Q$ in \bref{LQ}, $L_P^{\hat 0}$ in \bref{LP} and 
$\CF^{\underline{j}}$ in \bref{ninv2f}
have the SO(2,1) symmetry.
As the $H$ invariant Lagrangian we take the Lagrangian similar
 form as that in \cite{Towc}.
For a D1 case,
\bea
\CL~&=&~\frac{1}{2e}\left(\det G_{\mu\nu}+{\cal F}^{\underline{j}}
{\cal F}_{\underline{j}}~\right)
\label{solwz1}
\eea
with ${\cal F}^{i}$ given \bref{ninv2f}. 
The difference is that the U(1) potentials $A^{\underline i}$ are not 
fundamental worldvolume fields but given in terms of
new superspace coordinates in \bref{solwz1}. 
In \cite{Sakag} the same form of the 
Lagrangian is examined{
\footnote{
The SL(2,R) covariant algebra considered in the ref.\cite{Sakag}
 is not (graded)
Lie algebra since the structure constants depend on the dilaton and
axion explicitly.}
}
and is shown to have the kappa invariance.
Here we clarify the canonical structure and show how the U(1) degrees of
freedoms are 
incorporated in terms of new superspace coordinates.

In order to discuss it the explicit form of the Lagrangian is
not necessary we only assume the Lagrangian is described as a 
$H$ invariant function of form{\footnote{Lagrangians may depend
on the higher rank tensors also in case of $p>1$.}}
\bea
\CL&=&\CL(G_{\mu\nu},\CF^i).
\eea
The new variables $y^{\underline i}$ and $\xi^i$ enter in the Lagrangian 
only through $F^{\underline {i}}$ in ${\cal F}^{\underline {i}}$
\bea
F^{\underline {i}}_{\mu\nu}&=&\partial_{[\mu}A^{\underline {i}}_{\nu]}~
=~\partial_{[\mu}[
-\frac{1}{2}\{ y^{\underline {i}}\partial_{\nu]} x +
(\overline\xi^{\underline {i}}c\varrho^{\hat 0}+
 \overline\xi^{\hat{0}}c\varrho^{\underline {i}})
\partial_{\nu]}{\theta}\}].
\eea
The variation of the action with respect to $y^{\underline {i}}$,
$\xi^{\hat 0}$ and 
$\xi^{\underline i}$ gives the equations of motion, 
\bea
\pa_\mu(\frac{\pa L}{\pa \CF^{\underline {i}}_{\mu\nu}})~\pa_\nu x^m~=
\pa_\mu(\frac{\pa L}{\pa \CF^{\underline {i}}_{\mu\nu}})
c\varrho^{\underline {i}}~\pa_\nu \T~=~
\pa_\mu(\frac{\pa L}{\pa \CF^{\underline {i}}_{\mu\nu}})
c\varrho^{\hat 0}~\pa_\nu \T~=~0.
\eea    
If the induced metric is not singular the independent equations are
\bea
\pa_\mu(\frac{\pa L}{\pa \CF^{\underline{i}}_{\mu\nu}})~=~0.
\label{maxw}
\eea
It is the Maxwell equation, which would be obtained when $A^{\underline {i}}$ 
were independent world volume fields. 
It shows that $y^{\underline {i}}, \xi^0$ and $\xi^{\underline {i}}$ 
are not dynamically independent but
have same dynamical modes of U(1) potentials $A^{\underline {i}}$.
The U(1) gauge transformation of $A^{\underline{i}}$ is induced by that of
the new variables, for example, 
\bea
\D y^{\underline{i}m}&=&-2\pa_\mu\lam^{\underline{i}}G^{\mu\nu}\Pi_\nu^m,~~~~~
\D \xi^{\underline{i}}~=~2\pa_\mu\lam^{\underline{i}}G^{\mu\nu}
\slPi_\nu\T,~~~~~\to~~~~~
\D A_\mu^{\underline{i}}~=~\pa_\mu\lam^{\underline{i}}
\eea
where $\lambda^{\underline{i}}$ are U(1) gauge parameters.
\vskip 6mm

In the canonical formalism
the canonical variables are $x^m,~ \T,~ y^{\underline{i}}=(y^{\hat{1}},y^{\hat
{2}}),~ \xi^j=(\xi^{\hat{0}},\xi^{\hat{1}},\xi^{\hat{2}})$ and their canonical 
conjugates $~p_m,~\z,~p^y_{\underline {i}}=(p^y_{\hat{1}},p^y_{\hat{2}}),~
\pi_j^{\xi}=(\pi^\xi_{\hat{0}},\pi^\xi_{\hat{1}},\pi^\xi_{\hat{2}}) $ 
respectively.
Defining equations of canonical conjugate momenta for new variables
are 
\bea
p^y_{\underline{j}}&=&\frac{\partial {\cal L}}
{\partial {\dot y^{\underline{j}}}}
=-\frac{1}{2}(E_{\underline{j}})^a
\partial_a x,
\label{py}
\\
\pi^\xi_{\hat{0}}&=&\frac{\partial^r {\cal L}}{\partial \dot{\xi}^{\hat{0}}}
=-\frac{1}{2}(E_{\underline{j}})^a\partial_a \bar{\theta}c\varrho^
{\underline{j}},
\label{pix0}\\
\pi^\xi_{\underline{j}}&=&\frac{\partial^r {\cal L}}{\partial \dot{\xi}^
{\underline{j}}}
=-\frac{1}{2}(E_{\underline{j}})^a\partial_a \bar{\theta},
\label{pixi}
\eea 
where 
\bea
(E_{\underline{i}})^a&\equiv&\frac{\partial 
{\cal L}}{\partial {\cal F}^{\underline {i}}_{0a}
}(\dot{x},\dot{\theta},\dot{y},\dot{\xi}),~~~~~~~(a=1,...,p).
\eea
If we define the inverse of the spatial metric ${\bf G}^{ab}$ and
the following operator
\bea
~G_{ab}{\bf G}^{bc}=\delta_a^c~~,~~
\Upsilon_{mn}\equiv (\Pi_a)_m{\bf G}^{ab}(\Pi_b)_n~~,~~
\eea
then $\Upsilon$ is the projection operator to  
spatial tangential direction of a p-brane; 
~$\Upsilon_m^n \Upsilon_n^l=\Upsilon_m^{~l}$~~,~~$
(\Pi_a)^m\Upsilon_m^{~n} =(\Pi_a)^n~~,$~~rank$(\Upsilon_m^{~n})=p$.

Combining \bref{py} and \bref{pixi}
$~E^a$ is solved in terms of the canonical variables as
\bea
(E_{\underline{j}})^a=-2\left(p^y_{\underline{j}m}+\pi^\xi_{\underline{j}} 
\Gamma_m\theta\right) 
(\Pi_b)^m{\bf G}^{ba}~.\label{Epy}
\eea
Inserting \bref{Epy} back to \bref{py}, \bref{pix0} and \bref{pixi} we get 
first class constraints
\bea
\left\{\begin{array}{ccl}
\phi_{\underline{j}}^y&=&\left((p^y_{\underline{j}})^m+
\pi_{\underline{j}}^\xi\Gamma^m\theta\right) (\eta_{mn}-\Upsilon_{mn})=0,
\\
\phi_{\hat{0}}^\xi&=&\pi^\xi_{\hat{0}}-(p^y_{\underline{j}}+\pi^\xi_
{\underline{j}}\Gamma \theta)\cdot\Pi_b{\bf G}^{ba}\partial_a\bar{\theta}
(c\varrho^{\underline{j}})=0,
\\
\phi_{\underline{j}}^\xi&=&\pi^\xi_{\underline{j}}-(p^y_{\underline{j}}+
\pi^\xi_{\underline{j}}\Gamma \theta)\cdot\Pi_b{\bf G}^{ba}\partial_a\bar{\theta}
=0.
\\
\end{array}\right.
\label{pricon}\eea 
The first constraint restricts that the dynamical degrees of freedom of 
$p^y_{\underline{j}}$ is lying on the p-brane. 
The latter two tells that $\xi^i$'s are gauge degrees of freedom.
The second one is written using the third one as
\bea
\phi^\xi_j\varrho^j=\pi^\xi_{j}\varrho^{j}=0.
\label{Zgen}
\eea
This is a canonical realization of the irreducibility condition of $Z_{iA}$
in \bref{ZREL}, i.e. \bref{Zgen} is the generator of the gauge 
transformation \bref{xigauge}. It guarantees the (on-shell) closure of the
SO(2,1) covariant super-Poincar$\acute{\rm e}$ algebra $G$.
\vsej

It is possible to find a canonical transformation in which
$(E^a_{\underline{j}},A_a^{\underline{j}})$ are {\it new} (tilde) canonical 
pairs.
The generating function is
\bea
{\cal W}(q,\t p)&=&\int~\left(\right.\t p x~+~\t\z\T~+~
\t p^{ym}_{\underline{j}}(\eta_{mn}-\Upsilon_{mn})y^{\underline{j}n}~+~
\t\pi^\xi_{{j}}\xi^j
\nn\\
&+&E^a_{\underline{i}}
[
-\frac{1}{2}\{ y^{\underline {i}}\partial_{a} x +
(\overline\xi^{\underline {i}}c\varrho^{\hat 0}+
 \overline\xi^{\hat{0}}c\varrho^{\underline {i}})
\partial_{a}{\theta}\}]\left.\right)~,\label{cangen}
\eea
where tilde coordinates are $(\t x, \tilde{\theta}, \t y^{\underline{j}},
A_a^{\underline{i}},\t\xi^{{j}})$ and the conjugate momenta are 
$(\t p,\t\z,\t p^y_{\underline{j}},E^a_{\underline{i}},
\t\pi^\xi_{{j}})$.
$(\t y^{\underline{j}m}, \t p^y_{\underline{j}m})$ are 
$(10-p)\times 2$ independent canonical pairs. 
Using tilde variables the constraints \bref{pricon} take simple forms
\bea
\t p^y_{\underline{j}m}&=&\t\pi^\xi_{{j}}~=~0.
\label{pricon2}
\eea
The Gauss law constraints are obtained as the secondary constraints as
\bea
\pa_a E^a_{\underline{j}}&=&0.
\label{Gauss}\eea

\medskip

In the case of $p=1$ 
the D1-brane Lagrangian \bref{solwz1} gives following constraints
in addition to \bref{pricon}
in terms of the {\it original} variables $( y^{{j}},{\theta},
\xi^{{j}})$ and $( p^y_{j},\z, \pi^\xi_{{j}})$;
\bea
H&=&\frac12~[~(\hat{p}^y_{\hat{0}})^2~+~4(p^y_{\underline{j}})^2]~+~
(~\z~{\varrho}^0\hat{\varrho}~\T'~
+~\pi^\xi_{\hat{0}}~{\varrho}^0\hat{\varrho}~\xi^{\hat{0}'}~
+~\pi^\xi_{\underline{j}}~{\varrho}^0\hat{\varrho}~\xi^{\underline{j}'}~)
=~0,
\label{defHH}\\
T_1&=&\hat{p}^y_{\hat{0}}~ x'~+~
(~\z~\T'~+~\pi^\xi_{\hat{0}}~\xi^{\hat{0}'}~
+~\pi^\xi_{\underline{j}}~\xi^{\underline{j}'}~)
=~0,
\label{defTT}\\
F&=&\z~-~p^y_{\underline{i}}\cdot x'\frac{1}{(x')^2}
(\bar{\xi}^{\underline{i}'}c{\varrho}^0+
 \bar{\xi}^{\hat{0}'}~c\varrho^{\underline{i}})~
+~\Tb\Gamma\cdot
(\hat{p}^{y}_{\hat{0}}~+~\frac12(\Tb\G c\hat{\varrho}\T'))c{\varrho}^0
\nn\\
&&-~
( x'-\frac12\Tb\G c{\varrho}^0\T')\cdot\Tb\G c\hat{\varrho}~=~0,
\label{defFF}\eea
with 
\bea
\hat{p}^y_{\hat{0}}={p}^y_{\hat{0}}+({p}^y_{\underline{j}}\cdot 
x'\frac{1}{(x')^2}
y^{\underline{j}})'
~~,~~
\hat{\varrho}=-2{p}^y_{\underline{j}}\cdot x'\frac{1}{(x')^2}
\varrho^{\underline{j}}~~.
\eea
For a D-brane its static property $(x')^2\neq 0$ is assumed.
Performing the canonical transformation \bref{cangen},
above constraints \bref{defH}, \bref{defT} and \bref{defF} are
rewritten in terms tilde coordinates  
in addition to \bref{pricon2} and \bref{Gauss},
\bea
H&=&\frac12~[~\tilde{p}^2~+~({ E}^1_{\underline{j}})^2~(\t x')^2]~+~
\t\z~{\varrho}^0\varrho_E~\t\T'~=~0,
\label{defH}\\
T_1&=&\tilde{p}~\t x'~+~\t\z~\t\T'~=~0,
\label{defT}\\
F&=&\t\z~+~\t\Tb\Gamma\cdot(\t p~+~\frac12
(\t \Tb\G c\varrho_E\t \T'))c{\varrho}^0
\nn\\
&&-~
(\t x'-\frac12\t\Tb\G c{\varrho}^0\t\T')\cdot\t\Tb\G c\varrho_E~=~0,
\label{defF}\eea
where 
\bea
\varrho_E&\equiv&E^1_{\underline{j}}~\varrho^{\underline{j}}.
\eea
In the tilde canonical coordinates the extra variables $\t y^{\underline{j}}$ 
and $\t\xi^j$ 
and their conjugate momenta are decoupled and the set of constraints
\bref{defH}-\bref{defF} have the same forms as ones of 
\bref{solwz1} in which the U(1)
potentials are regarded as independent fields.
Comparing with \bref{defHH} and \bref{defH}, it can be seen that U(1) momenta 
contribute to $H$ in a same form as the brane momentum.
From the construction there exists the SO(2,1) symmetry which 
mixes brane coordinates and U(1) modes, i.e. they are coordinates of the enlargespace. 
Therefore the SO(2,1) covariant superalgebra with central extension 
explain geometrical origin not only of IIB SL(2,R) symmetry but also
of world volume U(1) modes.
  
\vsej


\section{Discussions}\par
\indent

In this paper we have proposed a SO(2,1)$\cong$SL(2,R) covariant IIB 
superalgebra with central extension $G$.
It contains SO(2,1) triplet momenta, doublet SUSY charges and
spin 3/2 central charges.
In order to describe Dp-brane systems the coset, 
$G$/( SO(2)$\times$Lorentz ), is considered. 
Our parametrization of the coset naturally leads to the brane coordinates; 
a singlet position $x^m$ and SO(2) doublet fermionic coordinates $\T^{A\A}$. 
The dilaton and axion also appear as the coordinates of the coset. 
The SUSY invariant two form doublet is constructed and 
the U(1) potentials acquire an interpretation in terms of  
coordinates in the enlarged space.  

The  SO(2,1)/SO(2) coset element $ g_N $ naturally produce SO(2,1) covariant 
coordinates; SO(2,1) doublet from SO(2) doublet and SO(2,1) triplet from 
SO(2) triplet. 
However in order to obtain the IIB SL(2,R) supergravity doublet 
$(B^{NS},B^R)$ from the SO(2) {\it vector} doublet
the  SO(2,1)/SO(2) coset element $K$ is not sufficient. 
In order to solve this puzzle 
we have introduced one more 
auxiliary field $\tilde{\varphi}$ in addition to the SO(2) gauge parameter 
$\varphi$. It gives an alternative representation of the coset SO(2,1)/SO(2).
We showed that in a gauge $ \tilde{\varphi}+{\varphi}=0$  
the SO(2,1)/SO(2) coset element coincides with that of SL(2,R)/SO(2) 
expressed in terms of the dilaton and the axion. 
In a similar context, Bars looked for the origin of auxiliary degrees of 
freedom in the extra space; 14 dimensions \cite{Bars}.

We performed a central extension of IIB superalgebras not only in a SO(2,1) 
covariant manner but it fits with the Siegel's mechanism \cite{Siegel,Green}.
An advantage of the central extension is that new bosonic coordinates 
$y^{\underline{j}}$ and new fermionic coordinates $\xi^{{j}}$ 
appear in supersymmetric combinations to form the simple Wess-Zumino terms.
Dynamical modes of the new coordinates are restricted only on a brane
as \bref{pricon}
and they play a role of worldvolume U(1) gauge potentials.
Therefore the worldvolume U(1) fields can be recognized as the coordinates 
of the enlarged superspace.
It suggests that the 2 dynamical modes of the worldvolume U(1) field
of D3-brane can be recognized as coordinates in the 10+2 dimensions 
\cite{Tsy}.  
The SO(2,1) covariant superalgebra may give us new insight of 
the geometrical origin of the IIB SL(2,R) duality from higher 
dimensional view point.

\vskip 6mm
{\bf Acknowledgements}\par
M.H. is partially supported by the Sasakawa Scientific Research Grant 
from the Japan Science Society.

\end{document}